\documentstyle[12pt,epsf]{article}

\date{\today}

\newcommand\putfig[3]{
   \vbox{
   \let\picnaturalsize=N
   \def\picsize{#3}
   \def\picfilename{#1}
   \ifx\nopictures Y\else{\ifx\epsfloaded Y\else\input epsf \fi
   \let\epsfloaded=Y
   \centerline{\ifx\picnaturalsize N\epsfxsize \picsize\fi
   \epsfbox{\picfilename}}}\fi
   \vspace{1.0cm}
   {\it #2}
   \vspace{1.5cm}
   }
}

\def\be{\begin{equation}}
\def\ee{\end{equation}}
\def\bear{\begin{eqnarray}}
\def\eear{\end{eqnarray}}
\def\nn{\nonumber}

\def\hlf{{{1\over 2}}}

\def\wdg{{\wedge}}                              % wedge product

\newcommand\inv[1]{{1\over{#1}}}

\newcommand\rep[1]{{\underline{\bf {#1}}}}      % representation
\newcommand\trr[2]{{\mbox{tr}_{#1}\{{#2}\}}}    % trace in a rep
\def\IZ{{\bf Z}}                                % Math Z
\newcommand\MR[1]{{{\bf R}^{#1}}}               % Real numbers
\newcommand\MS[1]{{{\bf S}^{#1}}}               % Circle, sphere,...
\newcommand\MT[1]{{{\bf T}^{#1}}}               % Torus
\newcommand\CP[1]{{{\bf CP}^{#1}}}              % CP
\newcommand\MFD[1]{{X_{#1}}}             % manifold
\newcommand\MSP[1]{{{\cal M}_{#1}}}             % moduli space
\def\a{{\alpha}}
\def\b{{\beta}}

\def\u{{\mu}}
\def\v{{\nu}}

\def\t{{\tau}}

\def\lam{{\lambda}}

\def\Hil{{\cal H}}                              % Hilbert space
\def\PF{{\cal Z}}                               % partition function
\newcommand\SUSY[1]{{{\cal N}= {#1}}}           % N=? SUSY

\def\kbl{{\cal K}}                              % canonical bundle
\def\nbl{{\cal N}}                              % normal bundle
\def\trv{{\cal O}}                              % trivial bundle

\newcommand\Bm[1]{{B^{(-)}_{#1}}}               % anti-self-dual 2-form

\def\npb#1#2#3{{\it Nucl.\ Phys.} {\bf B#1} (19#2) #3}

\def\prd#1#2#3{{\it Phys.\ Rev.} {\bf D#1} (19#2) #3}

\def\hepth#1{{\it hep-th/{#1}}}

\begin{document}
\begin{titlepage}
\titlepage
\rightline{PUPT-1634}
\rightline{hep-th/9607092}
%%% \rightline{DRAFT VERSION}
\rightline{July 11, 1996}
\vskip 1cm
\centerline {{\Large \bf Compactification of Tensionless String Theories}}
\vskip 1cm
\centerline {{\bf Ori J. Ganor}}
\vskip 0.5cm
\begin{center}
\em  origa@puhep1.princeton.edu\\
Department of Physics, Jadwin Hall, Princeton University\\
Princeton, NJ 08544, U. S. A.
\end{center}
\vskip 1cm
\abstract{
 We study compactifications of the $N=2$ 6D tensionless string
on various complex two-folds down to two-dimensions.  In the IR
limit they become non-trivial conformal field theories in 2D.
Using results of Vafa and Witten on the partition functions of
twisted Super-Yang-Mills theories, we can study the resulting CFT.
We also discuss the contribution of instantons made by wrapping
strings on 2-cycles of the complex two-fold.
}
\end{titlepage}

%%%%%%%%%%%%%%%%%%%%%%%%%%%%%%%%%%%%%%%%%%%%%%%%%%%%%%%%%%%%%%%%%%%%%%%%%%%%
%                      BEGINNING OF TEXT                                   %
%%%%%%%%%%%%%%%%%%%%%%%%%%%%%%%%%%%%%%%%%%%%%%%%%%%%%%%%%%%%%%%%%%%%%%%%%%%%

%==================================================================%
% Section (1): Introduction                                        %
%==================================================================%

\section{Introduction}

It has been suggested almost two years ago \cite{VWSCT} that
S-duality of $\SUSY{4}$ Yang-Mills theory can be better understood
as originating in a higher dimensional theory.
It was then shown in \cite{VerGLO} how the $\tau\rightarrow-\inv{\tau}$
duality of the pure   $U(1)$ gauge theory on a compact 4-manifold
is natural when the $U(1)$ theory is realized as a compactification
of an anti-self-dual free 2-form field $B_{\u\v}^{(-)}$ from
6D to 4D on a $\MT{2}$ whose complex structure is identified with
the complex coupling constant of $U(1)$.

A year ago, a new type of quantum theory,
which is probably not a field theory, was discovered in \cite{WitCOM}.
It was defined as the low-energy of type-IIB compactified on K3
with a 2-cycle that becomes very small.
The resulting low-energy description in 6D is a $\SUSY{2}$
theory of a massless tensor multiplet consisting of
an anti-self-dual 2-form $B_{\u\v}^{(-)}$ and 5 scalars (and their
super-partners).  The spectrum also includes a light BPS string 
which couples to $B_{\u\v}^{(-)}$.
Since the string is much lighter than the compactification scale
the theory can be considered on its own without taking into account
gravity and the other type-IIB modes.
The tension of the string is controlled by the VEVs of the scalars
and when the VEVs become zero the theory becomes scale invariant.

Theories with similar features have been encountered in various other
constructions.
In \cite{StrOPN} the 
world-volume theory of two parallel 5-branes in M-theory that
get very close was shown to contain the same kind of objects.
The connection between that construction and the previous one
was made explicit in \cite{WitFIV}.
The examples with $\SUSY{1}$ in 6D are richer.
In \cite{GanHan} it was argued that the phase discovered in \cite{DMW}
of a 5-brane in the bulk of M-theory on roughly $\MS{1}/Z_2\times K3$
can contain light strings with an $E_8$ current algebra on them.
An extensive description
of $\SUSY{1}$ tensionless string theories
that arise at phase transition points of heterotic vacua
on $K_3$ was given in \cite{SWCSD}.
More recently, examples of such theories with light stringy objects
have been given in 4D \cite{WitPMF,HanKle}. Their possible role
in exciting new phenomena in 4D \cite{WitPMF,HanKle}
and (of tensile strings)
in black-hole physics \cite{DVV1,DVV2,Mal,HanKle,DVV3} has been suggested.

All these constructions \footnote{The
construction of \cite{HanKle}
can probably also be realized as an orbifold of M-theory on $\MT{7}$
with intersecting 5-branes in the bulk,
which is probably dual to a theory on the same moduli space of
type-IIB on a Calabi-Yau as in \cite{GopMuk}.}
can be described in the framework of
 F-theory \cite{VafaFT,SWCSD,WitPMF,MVI,MVII} as
type-IIB vacua on a base manifold with a 2-cycle
that shrinks to zero.

In this paper we will study the (probably) simplest of those
new theories, i.e. the one with $\SUSY{2}$ in 6D.

For lack of a better name, we will refer to it as 
{\em S-theory}.
We will use that term for both the ``tensionless'' theory as well
as the tensile theory which differ from one another
by the values of the VEVs of the tensor-multiplet.

The question that we wish to ask is {\em ``what happens to
S-theory upon compactification?''}.

Compactifying on $\MT{2}$ was argued \cite{WitCOM} to give
an alternative construction of $\SUSY{4}$ $SU(2)$ Super-Yang-Mills.
The charged W-bosons, monopoles and dyons of SYM
become the strings of S-theory
wrapped on 1-cycles of the torus $\MT{2}$.

In the present paper we will discuss compactification
on a 4-manifold $\MFD{4}$. In order to study a large
class of $\MFD{4}$-s and to preserve part of the
supersymmetry we will have to perform a twist.
The resulting theories are non-trivial 2D CFTs in the IR limit.
Those turn out to have $(0,2)$ SUSY for general complex manifolds,
$(0,4)$ for K\"ahler manifolds and $(0,8)$ for K3.

We will extract information about the theories in three ways:
\begin{enumerate}
\item
2D gravitational anomalies.
\item
The partition functions.
\item
An instanton calculation.
\end{enumerate}

A calculation of the partition function has been made possible
due to the work of Vafa and Witten on twisted $\SUSY{4}$ Yang-Mills
\cite{VWSCT}. They found that in many cases the
partition function of the 4D theory on a 4-manifold in terms
of the Yang-Mills coupling constant
$\tau= {{i}\over {g^2}} + {{\theta}\over {2\pi}}$ equals
the partition function of a simple rational conformal field theory
where $\tau$ now is the complex structure of the 2D torus.
This surprising connection between Yang-Mills instantons and
RCFTs has been interpreted recently for the case of K3 
as the manifestation of type-IIA on K3 and heterotic on $\MT{4}$
duality which replaces 4-branes
with elementary heterotic string excitations \cite{VafIOD}.
In this paper, however,
 we will follow the original suggestion of \cite{VWSCT}
(and that of \cite{VerGLO}) to identify these RCFTs with the left-moving
part after compactification to 2D.

Recent developments in string theory \cite{PolDBR,WitBST}
allow us to give a low-energy description of the fields on the 
tensile strings of S-theory  and to estimate the contribution of
instantons made by wrapping the strings on 2-cycles of $\MFD{4}$
as in \cite{WitNON}.
We will calculate the condition for such an instanton to contribute
in the IR limit.

One of the questions we wish to explore is to what extent can 
S-theory be described as a tensor multiplet together with strings.
In particular, is it possible to describe the compactification
by first reducing the free tensor multiplet and then 
adding the contribution of the strings?

We do not have a systematic way to implement that procedure
but for the case of K3 we will compare the result to a $\IZ_2$
orbifold of a free tensor multiplet
and check that there are no obvious instanton corrections.
In the comparison we will find a discrepancy that we were not
fully able to explain.

The paper is organized as follows:
\begin{enumerate}
\item
Section (2): Discussion of the setting for the compactification
           and the twist.
\item
Section (3): The anomaly matching conditions that allow us to derive
the difference between left moving and right moving central
charges in 2D.
\item
Section (4): The free field computation -- ignoring the contribution
 of the strings.
\item
Section (5): We calculate the anomaly condition following \cite{WitNON}
 for an instanton to contribute to the effective IR action.
\item
Section (6): Application to various $\MFD{4}$-s.
\item
Section (7): Discussion.
\end{enumerate}

%==================================================================%
% Section (1): Compactification                                    %
%==================================================================%
\section{Compactification}

We wish to learn more about S-theory (``tensionless strings'')
 by compactifying it to 2D.
Compactifying the scale invariant S-theory
on a 4-manifold $\MFD{4}$ of finite size we expect
to find in 2D a theory whose scale is related to the size of $\MFD{4}$.
We note that in 2D the would-be VEVs of scalars are better interpreted as
target space coordinates so the 5 scalars of the 6D tensor multiplet
(i.e. the super-partners of $B_{\u\v}^{(-)}$) have 
``fluctuating VEVs'' after compactification.
Since the VEVs of the 5 scalars determine the tension of the string
of S-theory we see that the compactification ``probes'' the 
tensile strings as well. In other words, whether we compactify
the tensile or tensionless S-theory we end-up with one and the same
theory in 2D.

The next step would be to take the IR limit of the 2D theory,
or put differently -- to take the size of $\MFD{4}$ to zero.
In order to extract information on the resulting theory, we need
some kind of ``topological rigidity''.
This is given by the assumption that we can ``twist'' the supersymmetry
in the compact directions and if we use a special twist we have the
extra advantage of making contact with the results of Vafa and Witten
on twisted $\SUSY{4}$ Yang-Mills.

\subsection{Twisting}

Although it seems unlikely
that 6D tensionless strings are described by a field-theory,
it was suggested by E. Witten that the theory might have local
currents like the energy-momentum tensor.
In that case, we can consider twisting the theory to a topological
one.
In fact, the compactified tensionless string theory can be realized
as type-IIB compactified on an 8-manifold with a 4-manifold $\MFD{4}$
of $A_1$ singularities. The tensionless string theory becomes 
{\em automatically} twisted in this setting \cite{BSV,KMP}.

The R-symmetry group of the $\SUSY{2}$ theory in 6D is $Sp(2)$ and
the SUSY charges are in the $\rep{4}$ of $Sp(2)$.
Since the R-symmetry is part
of the super-conformal group we expect it to be a good symmetry
\cite{WitPRV}.

The procedure of twisting embeds (part of) the R-symmetry group in the
Lorenz group. We will take a subgroup $SU(2)\subset Sp(2)$ of the
R-symmetry. Under $SU(2)_R$ the super charges transform as
$2(\rep{2})$.

 Let us compactify on $\MT{2}\times \MFD{4}$, where $\MFD{4}$
is a complex manifold of complex dimension~2.
The 6D $\SUSY{2}$ charges are in the $\rep{4}\otimes \rep{4}$ of
$Sp(2)_R\times SO(5,1)$.
 Decomposing under $SO(1,1)\times SO(4)\subset SO(5,1)$ we have
$\rep{4} = \rep{2}_{+}\oplus \rep{2}'_{-}$.
we find that the super-charges are in 
\be
(\rep{2}_R\oplus\rep{2}_R)\otimes 
(\rep{2}_{+}\oplus \rep{2}'_{-})
\ee
The twist that was used in \cite{VWSCT}
replaces both $\rep{2}_R$-s with the $\rep{2}$ of the
Lorenz group $SO(4)$.\footnote{For 4D SYM
there is also another twist that was explored
in \cite{Marcus}, but I do not see how it can be implemented
in our setting.}
This leaves us with two covariantly constant super-charges
on $\MFD{4}$ which would in general give $(0,2)$ SUSY in 2D.

\subsection{The amount of SUSY in 2D}

In fact for K\"ahler manifolds the amount of SUSY is larger.
The super-charges are in the $\rep{3}+\rep{1}$ of $SO(4)$.
The $\rep{3}$ will correspond to anti-self-dual 2-forms
on $\MFD{4}$. Every such 2-form that is in addition covariantly
constant will give rise to 2 additional super-charges in 2D.
The K\"ahler class is always anti-self-dual
and covariantly constant, so for K\"ahler manifolds we always
have $(0,4)$ SUSY in 2D.
If in addition, $\MFD{4}$ is a K3 we have 2 more covariantly
constant anti-self-dual 2-forms corresponding to the nowhere zero
sections of the trivial canonical bundle. Thus, in this case
we find $(0,8)$ SUSY in 2D.

\subsection{Non trivial CFTs in the IR limit.}

The IR limit of the 2D theory is some $(0,2)$ CFT.
We will now claim that it is not trivial and in fact
its partition function is known.

Writing the Hilbert space of the 2D CFT that we obtain in the form
\be
\Hil = \sum_\a L_\a\otimes R_\a
\ee
where $R_\a$ is a right super-Verma module
and $L_\a$ is a left Verma module with no supersymmetry,
the partition function is
\be
\PF(q,\bar{q}) = \trr{\Hil}{(-)^{F_L + F_R} q^{L_0}\bar{q}^{\bar{L}_0}}
 = \sum_{\a'}\chi_{\a'}(q)
\label{pf}
\ee
where $\a'$ are those $\a$-s with a super-symmetric vacuum for $R_{\a'}$
and $\chi_\a(q)$ is the character of the left moving
$L_\a$ module (with $(-)^F$ if it has fermions).
This is very similar to the elliptic genus
of $\SUSY{2}$ theories \cite{WitELG}.

Now we can make contact with the results of \cite{VWSCT}.
They expressed the partition function of twisted
$\SUSY{4}$ Super-Yang-Mills on a 4-manifold as a sum of 
characters of 2D CFTs. It is tempting to identify these
characters with the characters $\chi_\a(q)$.

We can calculate  the ``functional integral'' of tensionless
strings on $\MT{2}\times \MFD{4}$ \footnote{Although
we have no known Feynman path integral, the association of a numerical
value to every manifold, or more precisely a section over the moduli space
of the 6D manifolds is more general and can be constructed from a Hilbert
space approach.}
\be
{{\mbox{S-theory}}
\over
{\MT{2} \times \MFD{4}}}
\ee
in two ways.
In the first way we use
\be
{{\mbox{S-theory}} \over {\MT{2}}}
\longrightarrow_{IR}
\,\,\,\,{\mbox{$\SUSY{4}$ Yang-Mills}}
\ee
and then use the results of \cite{VWSCT} for $\SUSY{4}$
twisted Yang-Mills compactified on $\MFD{4}$.
In the second way we use
\be
{{\mbox{Twisted S-theory}} \over {\MFD{4}}}
= {\mbox{CFT}} + ({\mbox{massive}})
\ee
Here all the massive states in 2D are in $\SUSY{2}$ multiplets
(as a result of the twisting) so that only the $\SUSY{2}$ vacuum
contributes to the partition function -- which is then
multiplied by the character of the right-moving CFT.
It is important that this result is independent of the ratio of
sizes of $\MFD{4}$ and $\MT{2}$ because each of the two ways
of calculating is in a different limit.
Vafa and Witten \cite{VWSCT} expressed the twisted $\SUSY{4}$
path integral as a sum of characters of 2D CFTs.
We see now that it is very plausible that those CFTs are
related to the left moving $L_\a$-s.

In the rest of the paper we will
try to get more information on the full CFT as follows:

\begin{itemize}
\item
We will calculate $c_L - c_R$, the difference of central
charges on the left and on the right.
\item
Starting with the tensile theory,
we will wrap a  non-critical
string around a 2-cycle of the manifold to obtain an instanton
and check the condition for it to contribute to the low-energy.
\end{itemize}

\subsection{A note on type-IIB compactifications to 2D}

It was shown in \cite{VWOLT} that type-IIA on $K3\times K3$
has a $B_{\u\v}$ tadpole because of interactions of the form
$\int B\wdg Y_8$ where $Y_8$ 
is an 8-form made of the curvature tensor.\footnote{I am grateful
to E. Witten for pointing this out to me.}
It was further explained in \cite{SVWCON} that in order
to cancel the resulting $B$ charge one needs to add 24 elementary
strings reduced on $K3\times K3$ and filling the entire uncompactified
2D.

What is the manifestation of that for type-IIB on $K3\times K3$
(which is the theory we are interested in because type-IIB on a singular
K3 and more generally type-IIB on an 8-manifold
with a 4-manifold of $A_1$ singularities contains S-theory)?

Compactifying further on $\MS{1}$ to  $0+1$ dimensions
type-IIA and type-IIB are T-dual. $B_{\u\v}$ becomes  the winding number
of elementary type-IIA strings and the $B_{\u\v}$ tadpole
becomes a shift of 24 units in the winding-number charge of the vacuum.
In type-IIB we therefore expect to find a shift of 24 units in 
the momentum of the vacuum around $\MS{1}$.
This is measured by a shift in $L_0-\bar{L}_0$ and indeed
the 2D reduction of the self-dual 4-form $B_4$ of type-IIB gives
rise to chiral bosons each having a shift of $(-\inv{24})$ in $L_0$.
Just like the $B\wdg Y_8$  term the $(-\inv{24})$ shift
is also a one-loop effect.

A similar phenomenon happens when we compactify S-theory on K3.
We will find in section (4) that there are chiral bosons in 2D
which bring about a shift of $24(-\inv{24}) = -1$ in $L_0-\bar{L_0}$.

%==================================================================%
% Section (3): Anomaly conditions                                  %
%==================================================================%
\section{Anomaly conditions}

We can determine $c_L-c_R$ by an argument similar
to `t Hooft anomaly matching.
Perturbing the theory to a tensile string, 
only the tensor multiplet remains in the IR.
However, the 2D gravitational anomaly $c_L-c_R$ should 
stay the same.

Let us see what the tensile theory would give.
The $B_{\u\v}^{(-)}$ is unaffected by the twist and gives
$b^+$ left-moving scalars and $b^-$ right-moving scalars,
where ($b^-$) $b^+$ is the number of (anti-) self-dual two-forms
on $\MFD{4}$.
For K\"ahler manifolds, those are given by
\be
(b^+, b^-) = (b^{1,1}-1, 1 + 2b^{0,2})
\ee
Before the twist, the fermions are of the form $\lam^{\a I}$ with $\a$
a spinor index in $\rep{4}$ of $SO(5,1)$ and $I$ an index in the
$\rep{4}$ of the R-symmetry group $Sp(2)$.

The spinor index decomposes under $SO(1,1)\times SO(4)\subset SO(1,5)$:
\be
\rep{4} = \rep{2}_{+} \oplus \rep{2}'_{-}
\ee

The next step is to restrict to the subgroup $SU(2)_R\subset Sp(2)$
of the R-symmetry. The $\rep{4}$ of $Sp(2)$ becomes $2(\rep{2})$.
Next we identify $SU(2)_R$ with the subgroup of $SO(4)$.
This converts the $\lam$ field into two
fields  each transforming
in
\be
(\rep{2}\otimes\rep{2})_{+} + (\rep{2}\otimes\rep{2}')_{-}
=\rep{3}_{+} + \rep{1}_{+} + \rep{4}_{-}
\ee
Here $\rep{3}$ is the anti-self-dual tensors of $SO(4)$.
The two fields are real.

The 5 scalars  are in the $\rep{5}$ of $Sp(2)$ which decomposes
as
\be
\rep{5} = \rep{3} + 2(\rep{1})
\ee
under $SU(2)\subset Sp(2)$. The $\rep{3}$ would give
$2b^{2,0}+1$ bosons.  Altogether we find
$2b^{2,0}+3$ left moving and $2b^{2,0}+3$ right moving bosons.

To sum up:
\vskip 0.5cm
\vbox{
\begin{tabular}{||c|c|c||}
\hline
6D field    &   Left                 &  Right   \\
\hline
$\Bm{MN}$   &   $b^{1,1}-1$ bosons &  $1 + 2b^{2,0}$ bosons \\
$\lam$      &                        &  $2(2b^{2,0}+2)$ fermions \\
$\phi$      &   $2b^{2,0}+3$ bosons  &  $2b^{2,0}+3$ bosons \\
\hline
\end{tabular}
\vskip 0.5cm
{\bf Table (1)}: The fields of the 6D tensor multiplet after the twist.
}
\vskip 0.5cm
Thus
\be
c_L-c_R = b^{1,1} - 4b^{2,0} - 4
\ee
For K3 we find $c_L - c_R = 12$.

If we go further and try to guess from the results of \cite{VWSCT}
that  $c_L(K3)=24$ we find $c_R = 12$.
Moreover, according to \cite{VWSCT} we might guess
 that blowing up a point on K3
adds $\Delta c_L = +1$. Substituting
we find the conjecture that $c_R=12$ is true for K3
and blow-ups of K3.

%==================================================================%
% Section (4): Free field computation                              %
%==================================================================%
\section{Free field computation}

Let us suppose, just for a moment, that the strings of S-theory
are very tensile and we can neglect them altogether.
This leaves us with only the free fields of the tensor multiplet
in 6D.
We recall from section (2) that in 2D all tensions are probed
because scalar VEVs fluctuate -- but we take the free field
case as a starting point.

The results of the twist operation are given in table (1).

We see, for example, that for K3 the free field computation gives
24 free  left moving bosons and 8 free right moving bosons
with $(0,8)$ free fermion super-partners.

The partition function of \cite{VWSCT} for K3 is given
by (\ref{pfvw}).
Let us go back to (\ref{pf}) and see if this is what
we get entirely from free-fields.

We will discuss 
the oscillator part and its relation to
24 free left-moving bosons later,
but let us first see 
what the compactification radii of the 24 bosons are and
why in (\ref{pfvw}) the momenta of the bosons (i.e. $p_L$)
do not contribute.

The scalars $\phi_i$ are uncompactified and so have a matching
condition $p_L = p_R$.

The radii of the modes from the anti-self-dual 2-form $B_{\u\v}^{(-)}$
have been determined in \cite{VerGLO}.
The requirement is that the 3-form field strength
obtained from $B_{\u\v}^{(-)}$ gives an integer multiple of $2\pi$
when integrated on 3-cycles. Here for a 3-cycle we take $\MS{1}$
(compactifying the spatial direction of 2D) times a 2-cycle
in $H^2(\MFD{4},\IZ)$.
It is interesting to note that in \cite{VerGLO} this requirement
was necessary to obtain compact $U(1)$ in 4D, but now we really
{\em need} this requirement already in 6D because there are
{\em physical} objects whose $B_{\u\v}^{(-)}$ charge is quantized!

Thus, the left moving and right moving bosons that come from
$B_{\u\v}^{(-)}$ live on a lattice of signature $(b^{1,1}-1, 2b^{2,0}+1)$
that corresponds to the embedding of self-dual and anti-self-dual
2-forms inside $H^2(\MFD{4},\IZ)$ \cite{VerGLO}.

The fermions are in the R-sector, so it seems at first sight that
there are no states in short $\SUSY{2}$ multiplets.
However, there is an extra $\IZ_2$ that we didn't yet take into account.

\subsection*{The extra $\IZ_2$ gauge symmetry}
To complete the free field computation we need to orbifold
by an extra $\IZ_2$ that reverses the sign of all the fields
in the tensor multiplet:
\be
\IZ_2: B_{\u\v}^{(-)} \leftrightarrow -B_{\u\v}^{(-)},\qquad
\phi^i\leftrightarrow -\phi^i,\qquad \lam\leftrightarrow -\lam
\ee
The necessity to include this $\IZ_2$ is best seen from the
description of \cite{StrOPN} of the theory of two 5-branes that
are close. The tensor multiplet that couples to the strings of
S-theory is the difference between the two free low-energy fields
on each 5-brane. The $\IZ_2$ global symmetry corresponds to
interchange of the two 5-branes. After compactification to 4D
it is related to the Weyl group of broken $SU(2)$ $\SUSY{4}$ Yang-Mills
and after compactification to 2D, as in our case, the $\IZ_2$
becomes a gauge symmetry and we need to orbifold by it.

Now we can see what are the short multiplets that enter into (\ref{pf}).
Since the $\IZ_2$ acts as $(-)^{F_R}$ on the fermion zero modes,
states with $p_R = 0$ and no right moving oscillators
will contribute to (\ref{pf}).

The restriction on the $\phi$-s sets $p_L = p_R = 0$.

For generic K3-s the $H^{1,1}$ cohomology
2-forms give non-integer numbers when integrated on 2-cycles
(because the $H^{2,0}$ cohomology mixes with $H^{1,1}$ to form
$H^2(\IZ)$) and in fact for generic K3-s no combination
of 19 self-dual 2-forms is in $H^2(\IZ)$.
Thus $p_R = 0$ forces $p_L = 0$ for the $B^{(-)}_{\u\v}$
modes as well.

For a blow-up of K3, on the other hand, the exceptional
divisor corresponds to an anti-self-dual integral $(1,1)$ class 
and thus the mode that comes from it is compactified on a radius
of $1/sqrt{2}$.
The $\sqrt{2}$ normalization is because, in the picture of \cite{StrOPN},
the strings are charged under the difference of two normalized 
$B_{\u\v}^{(-)}$-s, or in the construction of \cite{WitCOM}, the 2-cycle
that shrinks to zero has self-intersection $(-2)$ and not $(-1)$.
in general the radius is $1/(\sqrt{2}|(C\cdot C)|)$ for an extra cycle $C$.

Let us calculate the resulting partition function for K3.
In the untwisted sector there are 8 zero modes of the fermions on which
the $\IZ_2$ acts like the fermion number.
States with non-zero $(p_L, p_R)$ will be matched with $(-p_L,-p_R)$
and those two states have different fermion numbers so their
total contribution cancels.
States with zero momenta but an even number of left-moving oscillators
will have to have an even number of zero modes and come with
a $(+)$ sign while states with odd number have an odd number of zero
modes and come with a minus sign. Altogether we find

\be
{{2^3}\over{q\prod_{n=1}^\infty (1+q^n)^{24}}}
= 2^3{{\eta(q)^{24}}\over {\eta(q^2)^{24}}}
\ee
As for the twisted sectors, all oscillators have $\hlf$-integer
modes including the fermions. Thus there are no zero modes
but there are $2^{19+3}$ fixed points
and the twisted sector gives
\be
2^{19+3}\left(\inv{2q^{1/2}\prod_{n=1}^\infty (1-q^{n-\hlf})^{24}}
-\inv{2q^{1/2}\prod_{n=1}^\infty (1+q^{n-\hlf})^{24}}\right)
= 2^{21}{{\eta(q)^{24}}\over {\eta(q^{1/2})^{24}}}
+ 2^{21}{{\eta(q)^{24}}\over {\eta(-q^{1/2})^{24}}}
\ee

Defining
\be
G(q) = \inv{\eta^{24}}
\ee
we find altogether
\be
\PF =  \inv{G(q)} ( 2^3 G(q^2) + 2^{21} G(q^{1/2}) + 2^{21} G(-q^{1/2}))
\label{pfffz2}
\ee
We note that this is not completely modular invariant.
For modular invariance we should have given a different weight
to the twisted sectors since the modular invariant combination
is
\be
\PF =  \inv{G(q)} ( 2^3 G(q^2) + 2^{15} G(q^{1/2}) + 2^{15} G(-q^{1/2}))
\ee

This is the time to recall the results of Vafa and Witten \cite{VWSCT}:
\bear
G(q) &=& \inv{\eta^{24}} \nn\\
\PF_{SU(2)}(q=e^{i\tau}) &=& \inv{8} G(q^2) + \inv{4} G(q^{1/2})
                +\inv{4} G(-q^{1/2}) \nn\\
\PF_{SO(3)}(q=e^{i\tau}) &=& \inv{4} G(q^2) + 2^{21} G(q^{1/2})
                +2^{10} G(-q^{1/2}) \nn\\
\nn\\ &&
\label{pfvw}
\eear

There are several differences between (\ref{pfvw})
and (\ref{pfffz2}). First there is the overall factor of $G(q)$
that is missing in (\ref{pfffz2}) and then the relative
coefficient between the twisted and untwisted sectors in
(\ref{pfffz2}) does not agree with the coefficients of 
$G(q^{1/2})$ and $G(-q^{1/2})$ in either of the two formulas
(\ref{pfvw}).

Before we try to interpret the discrepancy between (\ref{pfvw})
and (\ref{pfffz2}) we have to ask two questions:
\begin{itemize}
\item
Is S-theory on $\MT{2}$ identical to $\SUSY{4}$  $SU(2)$ Yang-Mills
or are there topological restrictions on the gauge bundle? In
particular, should we pick $SU(2)$ gauge bundles or $SO(3)$ or
some combination of both?
\item
Do we expect the free field result above to be a good starting
point for obtaining the complete low-energy in 2D?
\end{itemize}

We start with the first question.
The authors of \cite{VWSCT} calculated the modular transformation
properties of their results (\ref{pfvw}) and found
\be
\PF_{SU(2)}(-\inv{\t}) = 
  2^{-12} \t^{-12} \PF_{SO(3)}
\ee
So we must start by asking whether it is possible that
S-theory on $\MT{2}\times K3$ is not modular invariant.
For 2D CFTs a modular anomaly usually appears when
we do not have a (manifestly covariant)
path-integral formula for the partition function
and have to resort to a trace in the Hilbert space (for example
a chiral boson away from the self-dual radius).
In this case, the partition function is not a number but a
{\em section} of a non-trivial line-bundle over the moduli
space of Riemann-surfaces. Thus, we could accept the modular
anomaly of S-theory if there is no (manifestly covariant)
path-integral for it only  some unknown Hilbert space.
This is not so surprising since $B_{\u\v}^{(-)}$ is chiral.

The next problem is the discrepancy in the relative coefficients
of $G(\pm q^{1/2})$ and $G(q^2)$.
The answer to that seems to be that we do not trust the twisted
sectors of the $\IZ_2$ orbifold. The fixed points are localized
near the region where the 5  scalars of the tensor multiplet are
zero. This is the region where the strings of S-theory are tensionless
and the free field approximation is the farthest from being trustworthy.

Nevertheless, it is amusing to note that all that is missing
seems to be a coefficient in the correct weight for
the twisted sector in order to get the $\PF_{SU(2)}$ result.

As for the question which gauge group we take, $SO(3)$ or $SU(2)$,
it seems that the answer lies in the twisted sectors on which we
have no information.

One possible answer might be that we actually get {\em both} $SU(2)$
and $SO(3)$ since we can move from one to the other by a ``change of
coordinates'' on the patch of the moduli space.

We just want to note one more thing concerning the $2^{-19}$
factor of $G(\pm q^{1/2})$.
The sum $G(q^{1/2}) + G(-q^{1/2})$ appeared in \cite{VWSCT} as
the contribution of  
gauge bundles of odd instanton numbers.
The $SU(2)$ result for even instanton numbers is \cite{VWSCT}:
\be
\PF_{SU(2)}^{(even)} = \inv{8}G(q^2)
\ee
and for odd instanton numbers \cite{VWSCT}:
\be
\PF_{SU(2)}^{(odd)} = \inv{4}G(q^{1/2}) + \inv{4} G(-q^{1/2})
\ee

We also note that from the above formula and the behaviour \cite{VWSCT}:
\be
\PF\sim \inv{q^2}
\ee
we can read off the momentum of the vacuum $L_0-\bar{L}_0 = -2$ if this
is indeed the partition function of a CFT.

Finally we wish to understand where the missing overall factor of $G(q)$
could possibly come from.

\subsection{The effect of the strings}
The ``spectrum'' of S-theory contains the tensor multiplet and
the strings. The free field computation neglects the strings
and now we wish to discuss what they can do.

To start with, let us note a few possibilities regarding the 
applicability of the free fields as a starting point.

First it might be that the free fields account for all the low-energy
fields but there are {\em generated interaction terms}.
In the next section we will check the instanton interaction terms
and find that, for K3, instantons made by a string wrapping an
analytic 2-cycle do not contribute to the low-energy.
The other possibility is that the free fields (and their possible
generated interactions) are indeed there but there are {\em more
low energy fields}. This is the case, for example, when $\MFD{4}$
has non-trivial $\pi_1$ and the tensionless strings can wind
around 1-cycles. 
We note that for compactification on $\MT{2}$ there are no 
modes of the ``tensionless strings'' other than those from
$\pi_1(\MT{2})$ that give rise to massless fields.
In our case $\pi_1(K3)=0$ so we do not expect more fields from 
such a mechanism either.
The third possibility is of course that the free field orbifold
is a bad starting point altogether.

Let us recall what the strings are capable of doing:
\begin{itemize}
\item
They can wrap on 2-cycles of the K3 to produce instantons in 2D.
This will be the subject of the next section.
\item
They can wrap on 1-cycles to produce more low-energy fields in 2D,
but there are no topologically non-trivial 1-cycles on K3.
\item
They {\em cannot}
 ``wrap'' on 0-cycles of K3, i.e. be reduced on K3 and stretch
on the entire 2D because the $B_{\u\v}^{(-)}$ charge will have nowhere
to go.
\end{itemize}

None of these options seems to explain the missing $1/\eta^{24}$ factor,
but let us speculate.
It seems that what we need are states that are points in K3
and carry after compactification one unit of $L_0$ each. Then
for $n$ such states the moduli space will be $K3^n/S_n$ and 
quantization as in \cite{VafIOD} will give the partition function
of 24 bosons. Such states cannot be the strings of S-theory
because of $B_{\u\v}^{(-)}$ charge conservation and because
they would then interact with the fields $\phi_i$ which set the
tension.

%==================================================================%
% Section (5): Instanton calculation                               %
%==================================================================%
\section{Instanton calculation}

In 2D scalars have no VEVs.
Because correlators do not fall off at large distances,
flat directions are interpreted as target space coordinates
and not as a moduli space.
Thus, the field $\sum_1^5\phi_i^2$ that specifies the tension
of the string does not have a VEV in 2D but fluctuates over the entire
region. Nevertheless, we can think of a region of field-space
for which the tension is large.
In that region the leading low-energy action will be
the free field theory with the fields found above and we intend
to look for possible instanton corrections.

There are, however, (at least) three possible flaws in the plan.
First, we have to be sure that there are no low-energy fields
other than those from the free-field computation.
In particular, we have to restrict to manifolds with $\pi_1(\MFD{4}) = 0$
since other-wise the strings that would wrap 1-cycles in $\MFD{4}$
would give new low-energy fields.
Second, the twisted sectors of the $\IZ_2$ orbifold from the previous
section are localized near the small tension regime and that sector
might be qualitatively different in the full theory.
Bearing that in mind we will proceed.

What is the effect of the non-critical string?

A 6D string can wrap around a 2-cycle of $\MFD{4}$.
Such an instanton can correct the effective action.

The situation is quite similar to that of \cite{WitNON}
where a 5-brane that is wrapped on a 6-cycle in M-theory
compactification to 3D could produce a super-potential.

There are however two important differences between that case
and ours.
First the geometry in our case is simpler because dimensions
are lower and second the supersymmetry structure is different.

It was argued in \cite{WitNON} that only special 6-cycles
(with arithmetic genus equal to 1) can create a super-potential.
In four dimensions, to find whether a super-potential can
be generated one has to count the number of zero modes of
the instanton configuration that are in
the $\rep{2}$ of the Lorenz group $SO(3,1)$ and subtract from
it the number of fermionic zero modes in the other representation
$\rep{2}'$.  An F-term can be generated if this difference is 2.
In 2D things are slightly different.
We have to count the difference between the number of fermionic
zero modes in the $+\hlf$ of $SO(1,1)$ and in the $-\hlf$ of $SO(1,1)$
(i.e. left moving and right-moving).
However, whereas in 4D one can combine two $\rep{2}$-s to a scalar
(and form an F-term Yukawa coupling), in 2D unequal numbers of
left and right moving fermions can only be coupled to derivatives
of scalars. Such terms will be irrelevant in the IR.
In fact, an unequal number of left-moving and right-moving
zero modes would seem to give a non Lorenz invariant contribution
to the action.

There is, however, another contribution to the chirality in 2D.
The instanton term for a string that wraps on a 2-cycle
$D\subset \MFD{4}$ behaves like:
\be
e^{-A_D V^{-1/2} \sqrt{\sum_{i=1}^5 \phi_i^2}} e^{-i\int_D B^{(-)}}
\ee
The first exponent is the tension of the string in terms of the
scalars ($A_D$ is the area of $D$ and $V$ is the volume of $\MFD{4}$).
The second term is the contribution of the interaction with the 2-form.
It is expressible as
\be
B^{(-)} = \sum_k \phi_L^k(x_0, x_1) \omega^{(-)}_k (x_2,\dots,x_5)
         +\sum_{k'} \phi_R^{k'}(x_0, x_1) \omega^{(+)}_{k'} (x_2,\dots,x_5)
\ee
where ($\omega^{(+)}$) $\omega^{(-)}$ are (anti) self-dual 2-forms
on $\MFD{4}$ and ($\phi_L$) $\phi_R$ are (left) right-moving bosons.

The chirality (i.e. $L_0 - \bar{L}_0$)  of the extra term 
$e^{-i\int_D B^{(-)}}$ is given by:
\be
\hlf\sum_k \left(\int_D\omega^{(-)}_k\right)^2
-\hlf\sum_{k'} \left(\int_D\omega^{(+)}_{k'}\right)^2
= (D\cdot D)
\ee
where $D\cdot D$ is the self-intersection of $D$.
(The $\sqrt{2}$ normalization is because, in the picture of \cite{StrOPN},
the strings are charged under the difference of two normalized 
$B_{\u\v}^{(-)}$-s, or in the construction of \cite{WitCOM}, the 2-cycle
that shrinks to zero has self-intersection $(-2)$ and not $(-1)$.)

We will see in the next subsection that $-2(D\cdot D)$ is exactly
the difference between right moving and left-moving zero modes.
So, counting $\hlf$ for each fermion  shows that the total $L_0-\bar{L}_0$
is zero, as it should.

\subsection{The low-energy fields on the string}

We will consider the case of a string wrapped on an analytic 2-cycle
$D\subset\MFD{4}$.

Let us determine first the low-energy fields on the non-critical string
after the twist.
The effective low-energy on the (tensile) anti-self-dual string
is the dimensional reduction of $\SUSY{1}$ 6D $U(1)$ SYM \cite{WitBST}
since it is derived from a wrapped 3-brane in type-IIB.
Under 
\be
Sp(2)_R\times SO(4)\times SO(1,1),
\ee
where the $SO(4)$ is the transverse Lorenz group and $SO(1,1)$
is the world-sheet, the fields transform as follows:
\vskip 0.5cm
\begin{tabular}{|cc|}
\hline
2D field    &   Representation   \\
\hline
$A_a$       &   $(\rep{1},\rep{1})_{+1} \oplus
                (\rep{1},\rep{1})_{-1}$ \\
$\lam$      &   $(\rep{4},\rep{2})_{+\hlf}$ \\
$\phi$      &   $(\rep{1},\rep{4})_{0}$  \\
\hline
\end{tabular}
\vskip 0.5cm
where $A_a$ is the world-sheet gauge field, $\lam$ are the fermions
and $\phi$ represent transverse oscillations of the world-sheet.

\subsection{Wrapping and Twisting.}
Now we embed the string in the compact 4 dimensions.
Let's call the non-compact directions $0,1$. The string is in
the compact $2,3$ directions and the other compact directions
are $4,5$.
What are the quantum numbers under
\be
SO(2)_N \times SO(2)_K \times SO(1,1)
\ee
where $SO(2)_N$ is rotations in the $4,5$ directions, $SO(2)_K$
is in $2,3$ and $SO(1,1)$ is in $0,1$?

The twist replaces the $\rep{4}$ of $Sp(2)$ with $2(\rep{2})$
of $SO(4)$ in the $2,3,4,5$ directions, so 
\be
\rep{4} \longrightarrow
2 (+\hlf,+\hlf,0) \oplus 2 (-\hlf,-\hlf,0) 
\ee
Also, the $\rep{2}$ of $SO(4)$ becomes
\be
\rep{2} \longrightarrow
(+\hlf,0,+\hlf) \oplus (-\hlf,0,-\hlf) 
\ee
(recalling that now the four directions orthogonal to the string
are $0,1,4,5$).
and the $+\hlf$ in the table above becomes
\be
()_{+\hlf} \longrightarrow (0,+\hlf,0)
\ee

The twist replaces $\lam$ with anti-commuting world-sheet fields in the
\be
\rep{4}\otimes\rep{2}\otimes (\cdots)_{+\hlf} \longrightarrow
2(+1,+1,+\hlf) \oplus 2(0,+1,-\hlf) \oplus 2(0,0,+\hlf)
\oplus 2 (-1,0,-\hlf)
\ee

Now we wrap around a 2-cycle $D$.
Let the canonical bundle of $D$ be $\kbl$ and the normal bundle
of $D$ in $\MFD{4}$ be $\nbl$.
So, in terms of bundles
we have right-moving fermions in
\be
2 \kbl\otimes \nbl + 2 \trv
\ee
and left-movers in
\be
2\kbl\oplus 2\nbl^{-1}
\ee
where $\trv$ is the trivial bundle.
We see that we get two generic right-moving zero modes as we should.
Those are the super-partners of translations \cite{WitNON}.
We also see that for K3, $\kbl\otimes\nbl$ is trivial and we 
get two more generic zero modes.

The difference between left and right moving zero modes is 
\be
2\{h^0 (\kbl\otimes\nbl) -h^0 (\kbl) - h^0 (\nbl^{-1})\}
\ee
Using the Riemann-Roch theorem this becomes
\be
2(c_1(\nbl) - 1) = -2(D\cdot D) - 2
\label{anco}
\ee
If we add to this the 2 generic zero modes from $\trv$
we find that this cancels the chirality of
$e^{-\int_D B^{(-)}}$ exactly.

\subsection{Conditions on 2-cycles}
If $-2(D\cdot D)-2 < 0$ it means that there are more left-moving
zero modes than right-moving ones (other than the 2 generic ones
from $\trv$). Since there are no left moving fermions the
instanton term will vanish in this case.
If $-2(D\cdot D)-2 = 0$ the resulting instanton term looks like
\be
e^{-A_D V^{-1/2} \sqrt{\sum_{i=1}^5 \phi_i^2}}\,\,\left(
e^{-i (\a \phi_R + \b \phi_L)} \psi_R^1\psi_R^2
\right)
\ee
where $\psi_R^i$ are right moving fermions and $\phi_R$ and $\phi_L$
are (right and left) chiral bosons. $\a,\b$ are some coefficients.
If $\a$ vanishes, we find (by a naive count of dimensions)
an  operator of dimension $(L_0=1,\bar{L}_0=1)$
in the large $\phi_i$ region.
If $\a$ does not vanish
or if $-2(D\cdot D)-2 > 0$ in which case there are more right moving
fermions, the instanton term will be irrelevant in the IR.
The condition that we found is thus $D\cdot D = -1$.
It also means that the cycle $D$ is isolated since $D\cdot D$ is also
the first Chern class of the normal bundle.

On a 2-cycle $D$ inside a K3, $\nbl_D=-\kbl_D$ and the anomaly
condition (\ref{anco}) implies $2g-2 = c_1(\kbl_D)  = -1$
which is impossible.
So no cycle meets the conditions
and it is plausible to expect no instanton correction.
An analytic cycle $D\subset \CP{2}$ always has positive intersection
and doesn't meet the condition either.
An exceptional divisor from a blown-up point, on the other hand,
is the generic example for a cycle that does meet the condition.

It is a bit puzzling however that this is the case because
the results of \cite{VWSCT} for a blow-up are essentially 
free-field results. On the other hand the result for $\CP{2}$
of \cite{VWSCT} have a holomorphic anomaly, i.e. are not a function
of $q$ but of $q$ and $\bar{q}$. It seems that an instanton might
cause such an effect because an irregular behaviour in the
spectrum of $\phi_i$ but we found that (at least for analytic cycles)
the instanton term vanishes.

Finally we note that
compactification on $\MT{2}\times \Sigma_g$,
a torus times a Riemann surface $\Sigma_g$
of genus $g$, can be studied from the results of \cite{BJSV,HMS}.
They discussed
compactification of partially twisted $\SUSY{4}$ on $\Sigma_g$
with non-trivial `t Hooft flux
and found a sigma model with Hyper-K\"ahler 
target space $\MSP{6g-6}$ of solutions
to a Hitchin system of equations.

This is an example with $\pi_1(\MFD{4})\ne 0$ where the extra fields
come from strings wound on the 2g cycles of $\Sigma_g$.

Compactifications of S-theory on $\Sigma_g$ might also be studied
in a setting similar to that of \cite{KLMVW} of 5-branes wrapped on 
$\MR{3,1}\times \Sigma_g$.

\section{Discussion}

Out of the new quantum theories M,F and S the latter which is
probably the simplest was the least explored.
Like M-theory and F-theory, the description of S-theory 
in its uncompactified form is unknown but compactifications
of S-theory can be described in a more conventional way in the
low-energy.

In this paper we were interested in the compactification down to 2D.
The identification of S-theory on $\MT{2}$ as $SU(2)$ $\SUSY{4}$ SYM
and the results of Vafa and Witten on twisted SYM partition functions
made it possible to study the 2D IR limit.
Furthermore, the low-energy description of the tensile strings
of S-theory, which was derived from the D-brane description and
the techniques developed in \cite{WitNON} enabled us to study
the contribution of strings wrapped on 2-cycles to the 2D low-energy.

We compared the results of Vafa and Witten for K3 to the contribution
of (a $\IZ_2$ orbifold of) the 6D tensor multiplet and 
found a missing $1/\eta^{24}$ factor which remained a puzzle.

We also found that anomaly conditions like in \cite{WitNON} seem 
to prevent an instanton contribution from wrapped strings for K3.

This paper dealt with what might be called
type-II S-theory (i.e. $\SUSY{2}$ in 6D). There are other
types as well. The one related to small $E_8$ instantons
might be called {\em heterotic} S-theory and it is interesting
to compactify that theory as well. This  time we have
to specify an $E_8$ gauge field  background  in addition to
the 4-manifold. It seems also natural to ask what are
orbifolds of S-theory? Are orientifolds possible? Is there
a type-I S-theory as well?

 Finally we come to the hardest question:  what did we learn about
the microscopic description of S-theory.
Even from the compactification on $\MT{2}$ we see
that the result can be described with only a finite number
of fields in the low energy. This is a bit mysterious
from the point of view of tensionless strings with an infinite
number of states becoming massless (see  e.g. \cite{LizSpa}).
Although this is not exactly a contradiction because most
of the string spectrum has width of order $1/\lam_{st}$ and
now $\lam_{st}\sim 1$ but still it is hard to see how
a conventional world-sheet approach could reproduce that.
The $\SUSY{2}$ strings \cite{OVgeom,OVhete}
which have recently been suggested
to play a fundamental role \cite{KutMar,KMO} seem to exhibit 
a better behaviour and in particular do not have the unwanted
tower of states but their critical dimension is not 6
and it is still hard to see how any world-sheet approximation could
be a good starting point when $\lam_{st}\sim 1$.
As was recalled in \cite{SchSDS} the Green-Schwarz action is
{\em classically} defined in $D=6$. There is a good S-theory
reason for that! It is the {\em low-energy} description of a tensile
string of S-theory. However
a low-energy approximation cannot be extrapolated to
a microscopic theory \cite{WitLEC}.

\section*{Acknowledgments}
I wish to thank M. Flohr, D.J. Gross, A. Hanany, E. Silverstein
and E. Witten for very helpful discussions.

\end{document}